\documentclass[10pt,english,oneside,twocolumn,a4paper]{article}
\usepackage{graphicx}
\usepackage[utf8]{inputenc}
\usepackage{times}
\fontfamily{ptm}\selectfont
\usepackage{tabularx}
\usepackage{ragged2e}
\usepackage[singlelinecheck=false]{caption}
\usepackage[T1]{fontenc}
\usepackage[backend=bibtex,doi=false,eprint=false,url=false,isbn=false,firstinits=true,sorting=none,style=numeric-comp]{biblatex}
\usepackage{amsmath}
\usepackage{amssymb}

\usepackage{babel}
\RequirePackage[blocks]{authblk}
\makeatletter
\let\ps@plain\ps@empty
\def\@xivpt{14pt}
\def\@sect#1#2#3#4#5#6[#7]#8{%
  \ifnum #2<2
    \null\par\vskip-15pt
  \fi
  \ifnum #2>\c@secnumdepth 
    \let\@svsec\@empty
  \else
    \refstepcounter{#1}%
    \protected@edef\@svsec{%
      \ifnum #2<4
        \hb@xt@10mm{\csname the#1\endcsname}\relax
      \else
        \hb@xt@12mm{\csname the#1\endcsname}\relax
      \fi}%
  \fi
  \@tempskipa #5\relax
  \ifdim \@tempskipa>\z@
    \begingroup
      #6{%
        \@hangfrom{\hskip #3\relax\@svsec}%
          \interlinepenalty \@M #8\@@par}%
    \endgroup
    \csname #1mark\endcsname{#7}%
    \addcontentsline{toc}{#1}{%
      \ifnum #2>\c@secnumdepth \else  
        \protect\numberline{\csname the#1\endcsname}%
      \fi 
      #7}%
  \else
    \def\@svsechd{%
      #6{\hskip #3\relax
      \@svsec #8}%
      \csname #1mark\endcsname{#7}%
      \addcontentsline{toc}{#1}{%
        \ifnum #2>\c@secnumdepth \else
          \protect\numberline{\csname the#1\endcsname}%
        \fi
        #7}}%
  \fi
  \@xsect{#5}}
\renewcommand\LARGE{\@setfontsize\LARGE{16}{20}}
\def\abstract#1{\def\@abstract{#1}}
\def\abstractEn#1{\def\@abstractEn{#1}}
\def\titleEn#1{\def\@titleEn{#1}}
\def\@maketitle{%
  \newpage
  \null
  \let \footnote \thanks
    {\LARGE\bfseries\RaggedRight \@title \par}%
    \vskip 1\baselineskip%
    {\normalsize
      \@author\par}%
    \vskip 2\baselineskip%
    {\section*{Abstract}
      \@abstract}%
  \par
  \vskip 3\baselineskip}

\renewcommand\section{\@startsection {section}{1}{\z@}%
                                   {-3.5ex \@plus -1ex \@minus -.2ex}%
                                   {\baselineskip}%
                                   {\normalfont\Large\bfseries\RaggedRight}}
\renewcommand\subsection{\@startsection{subsection}{2}{\z@}%
                                     {\baselineskip}%
                                     {1ex}%
                                     {\normalfont\large\bfseries\RaggedRight}}
\renewcommand\subsubsection{\@startsection{subsubsection}{3}{\z@}%
                                     {1\baselineskip}%
                                     {3bp}%
                                     {\normalfont\normalsize\bfseries\RaggedRight}}
\renewcommand\paragraph{\@startsection{paragraph}{4}{\z@}%
                                    {1\baselineskip\@plus1ex \@minus.2ex}%
                                    {3bp}%
                                    {\normalfont\normalsize\RaggedRight}}
\renewcommand\subparagraph{\@startsection{subparagraph}{5}{\parindent}%
                                       {3.25ex \@plus1ex \@minus .2ex}%
                                       {-1em}%
                                      {\normalfont\normalsize\bfseries\RaggedRight}}
\parindent\p@
\parskip\z@
\DeclareCaptionLabelSeparator{enskip}{\enskip}
\makeatother

\title{Quantum-Inspired Robust and Scalable SAR Object Classification}

\author[a,d]{Maximilian Scharf}
\author[a]{Marco Trenti}
\author[b]{Felix Bock}
\author[b]{Padraig Davidson}
\author[b]{Tobias Brosch}
\author[c]{Benjamin Rodrigues de Miranda}
\author[c]{Sigurd Huber}
\author[a]{Timo Felser}
\affil[a]{Tensor AI Solutions GmbH, Max-Rauth-Str. 16, Pfaffenhofen an der Roth, Germany}
\affil[b]{Hensoldt Sensors GmbH, Wörthstraße 85, Ulm, Germany}
\affil[c]{German Aerospace Center (DLR), Microwaves and Radar Institute, Oberpfaffenhofen, Germany}
\affil[d]{Ulm University, Institute for Complex Quantum Systems, Ulm, Germany}

\abstract{
SAR image classification naturally has to deal with huge noise and a high dynamic range
particularly requiring robust classification models. Additionally, the deployment of these models on edge devices, such as drones and military aircraft, requires a careful balance between model size and classification accuracy.
This study explores the potential of tensor networks to meet these robustness requirements, specifically evaluating their resilience to data poisoning.
Unlike previous works that concentrated on conventional neural networks for SAR object detection, this research focuses on the robustness and model reduction capabilities of tensor networks in object classification.
Our findings indicate that tensor networks are adept at addressing both the challenges of robustness and the need for model efficiency, thereby contributing valuable insights to the ongoing discourse in radar applications and deep learning methodologies in general.
}

\bibliography{library}

\begin{document}

\maketitle

\section{Introduction}

The success of deep learning
\cite{CiresanEtAlSchmidhuber2010,GravesSchmidhuber2009,KrizhevskySutskeverHinton2014,LeCunEtAlJackel1989,NeumannBrosch2020}
made it also popular for radar applications such as micro-Doppler classification \cite{NeumannBrosch2020, 7485236, 9673798, 8904525, AMIRI2020102702} or object detection in synthetic aperture radar (SAR) images
\cite{BroschNeuman2021, SchwaigerEtAlBrosch2022, Almalioglu_2025, 10911433, 2020AGUFMG004.0022W, PULELLA2024312}.
The huge dynamic range and inherent noisy nature of SAR images require a particular focus on the topic of robustness.
Furthermore, deployment on edge devices like drones or fighters necessitates a reduction in model size while maintaining a
high classification rate.
Here, we investigate, how tensor networks (TN) can solve the robustness requirements.
We evaluate the robustness to data poisoning using the popular MSTAR SAR classification task
\cite{KeydelLeeMoore1996,RossEtAlBryant1998}. Its moderate size while having a huge target variability, lends itself
perfectly to assess the various aspects of model reduction and robustness.
We do not focus on the task of SAR image formation (see \cite{CruzEtAlDuarte2022} for a recent overview), but on the
aspects of robustness and model reduction of TN to complement our recent investigations \cite{BroschNeuman2021,SchwaigerEtAlBrosch2022}. We demonstrate, that TN are well suited to address both challenges.

\section{Tensor Networks}

Tensor networks are a mathematical tool to deal with high dimensional data structures that first became popular in the study of quantum many-body systems \cite{Schollwoeck2011, SimoneBook}. Recently, they have been applied to other fields such as optimization \cite{Cavinato2021} and data science, in particular machine learning \cite{Stoudenmire2017, Felser2021}, as an alternative to neural networks. In this section, we briefly introduce the basic notions of TN and how they can be employed for supervised learning, followed by an information theoretical perspective of TN to illustrate some of their advantages over neural networks.

\subsection{Formal definition}

A tensor network is a tuple $(\mathcal{T}, \mathcal{V})$ of a set of tensors $\mathcal{T}=\{T_n\}_{n=1}^{N}$ and a set of links $\mathcal{V}=\{v_m\}_{m=1}^{M}$. 
Each tensor $T_{n}\in\mathbb{C}^{d_1\times d_2 \times ... \times d_{r_n}}$ has a rank $r_n$ and a set of indices $I_n=\{i_1,...,i_{r_n}\}$ with dimensions $d_1, ..., d_{r_n}$, respectively\footnote{While general TN are defined with complex tensors, especially in the context of quantum physics, we will only use real-valued TN with $T_{n}\in\mathbb{R}^{d_1\times d_2 \times ... \times d_{r_n}}$ in this application.}. 
Each link $v_m$ is an index that is shared by exactly two tensors $T_i$, $T_j$, such that $v_m=I_i\cap I_j$ and $v_m\notin I_k$ for $k\neq i,j$. 
Each link can also be interpreted as the edge of a graph, where the nodes are the tensors and the edges $e_m=\{T_i, T_j\}$ are given by the tensors that share the index $v_m$. Introducing the convention that indices that appear twice are implicitly summed over, each link also defines a tensor contraction. By summing over all links $v_m$, one can contract the whole tensor network, to obtain a single tensor $T_I=\sum_{v_1}...\sum_{v_m}(T_1)_{I_1}...(T_n)_{I_n}$, that consists of all the indices $I=\bigcup_{n=1}^{N}I_n\setminus \mathcal{V}$ that were not contracted and has therefore rank $r=|I|$. 
The tensor network $(\mathcal{T}, \mathcal{V})$ encodes the same information as the tensor $T$, but possibly with fewer parameters. The crucial degrees of freedom in a TN are the dimensions $\chi_m=\dim v_m$ of the links, that control the number of parameters needed to represent the TN. If these so-called \textit{bond dimensions} are bounded and not exponentially large with respect to the rank $r$, the TN is essentially a low-rank decomposition of the original tensor and requires fewer parameters to store the same amount of information.

In the context of quantum physics, the tensor $T$ typically encodes the state of a many-body quantum system. This state is given by a vector in a Hilbert space that can be represented as a tensor, where each index corresponds to a particle. 
The popularity of TN in this field arises from their ability to accurately represent ground states of many-body systems with gapped local Hamiltonians \cite{Hastings2006}, requiring exponentially fewer parameters than the full state vector. This allows a numerical representation which facilitates various computations for large systems which would otherwise be impossible.

\subsection{Tensor Network geometries}

\begin{figure}
  \includegraphics[width=\linewidth]{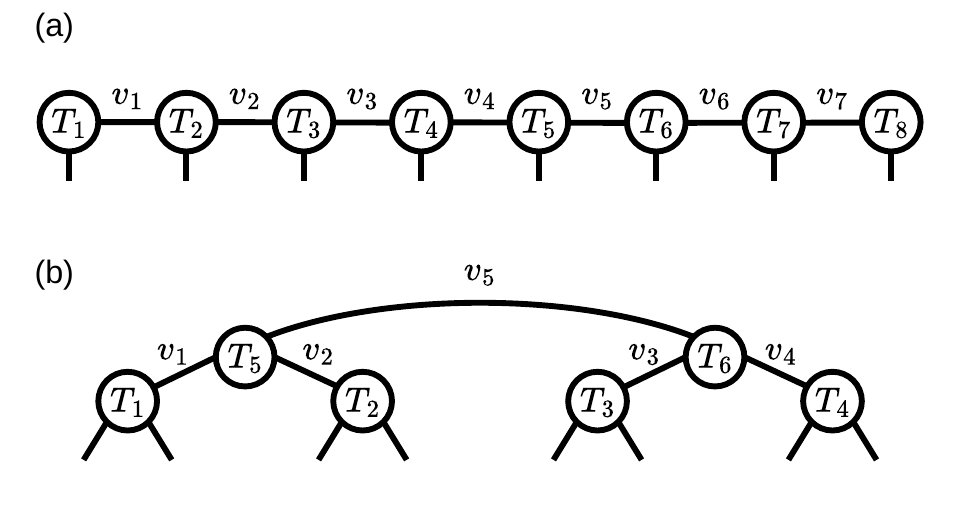}
  \caption{Example of the graph of an MPS (a) and a TTN (b) encoding a rank-8 tensor. The open edges correspond to the indices that are not contracted.}\label{fig:geometries}
\end{figure}

Depending on the internal structure of a tensor's information content, different graph structures of the tensor network yield better compression rates than others. The most prominent TN architecture is the class of matrix product states (MPS) \cite{Schollwoeck2011}, also referred to as tensor trains (TT), which is based on a linear graph. The MPS decomposition of a tensor $T$ with rank $r$ is given by
\begin{equation}
    T^{i_1...i_r}=(T_{1})_{v_1}^{i_1}(T_{2})_{v_1v_2}^{i_2}(T_{3})_{v_2v_3}^{i_3}\cdots\:(T_{r})_{v_{r-1}}^{i_{r}}\:,
\end{equation}
where indices that appear twice, the links, are implicitly summed over\footnote{Note that the indices $i_1, ..., i_r$ are written as upper indices only for the sake of clarity, as we are working in Euclidean space where there is no distinction between co- and contravariant space.}. The MPS/TT formalism is powerful, but limited to a certain class of high rank tensors that can be described efficiently, i.e., with bounded bond dimension. In the context of quantum physics, MPS/TT are well suited for 1D systems with local correlations, but fail to describe states of higher dimensional lattices. A generalization of the MPS format is the tree tensor network (TTN), also known as the hierarchical Tucker decomposition (HT), which is based on a binary tree graph. In \textbf{Figure~\ref{fig:geometries}}, the graphs of an MPS and a TTN are depicted for a small example. This graphical representation is identical to the Penrose notation of the tensor contractions and provides a simple but powerful way of depicting TN. Both MPS and TTN are loop-free TN, as their graph does not contain closed loops. TN geometries that do contain closed loops \cite{Evenbly2009, Felser2021a, Verstraete2006} can represent a larger class of states efficiently (with bounded bond dimension), but their usage in practice remains challenging, which is why we focus on the TTN in this work as the most flexible loop-free TN.

\subsection{Supervised Learning with Tensor Networks}
Supervised learning with tensor networks is based on the decision function
\begin{equation}
    f(\boldsymbol{x})=W\Phi(\boldsymbol{x})\, ,
\end{equation}
for a sample $\boldsymbol{x}\in X\subseteq\mathbb{R}^n$, where $\Phi: X \rightarrow \mathbb{R}^{d_1\times...\times d_n}$ is the feature map and $W\in\mathbb{R}^{d_1\times...\times d_n\times d_l}$ the weight tensor, with $d_l$ being the number of labels. When the two tensors are contracted, the final result is a vector of dimension $d_l$. Similar to kernel methods, non-linear decision boundaries can be achieved through the feature map $\Phi$ that maps the data to an exponentially higher dimensional space. While classical kernel methods rely on the kernel-trick to deal with the high dimensionality, the core idea of tensor network machine learning (TNML) is to encode both $\Phi(\boldsymbol{x})$ and $W$ as tensor networks \cite{Stoudenmire2017}. Starting with the feature map, it is typically defined as a tensor product
\begin{equation}
    \Phi(\boldsymbol{x}) = \phi_1(x_1)\otimes\phi_2(x_2)\otimes\cdots\otimes\phi_n(x_n)
\end{equation}
of local feature maps $\phi_i:\mathbb{R}\rightarrow\mathcal{H}_i$ that map a single feature $x_i$ to a vector of a local Hilbert space $\mathcal{H}_i$. A prominent choice for the local feature map is the \textit{spin map} $\phi(x_i)=(\cos(x_i\pi/2),\: \sin(x_i\pi/2))^{\textrm{T}}$ with $x_i\in[0,1]$, that maps each feature to a two-dimensional vector that can be interpreted as the state vector of a spin-$1/2$ particle in a Hilbert space $\mathcal{H}_i$ with dimension two. In this setting, one can not only interpret the input data as the state of a many-body-quantum system, but also the weight tensor $W$. This allows us to analyze the weight tensor, i.e., the model itself, with tools from quantum information theory, which will be discussed in more detail in the next section.

In the above definition, the weight tensor $W$ is exponentially large in the number of features. In order to be able to represent it numerically and optimize it, we encode it directly as a tensor network $(\mathcal{W, V})$ with tuneable bond dimensions, without ever having to construct the full tensor. This introduces bias, but enables us to use variations of efficient optimization algorithms for quantum systems like the density matrix renormalization group (DMRG) algorithm~\cite{Schollwoeck2011} or Riemannian optimization \cite{Willner2025}. Furthermore, the bond dimensions of the TN are directly related to the expressivity of the model, and can be tuned to avoid overfitting. This framework has been successfully applied to standard computer vision datasets such as the MNIST or fashion-MNIST, achieving state-of-the-art results \cite{Glasser2019, Stoudenmire2018, Stoudenmire2017}.

\subsection{Information theoretical perspective}
The interpretation of a TN as a quantum mechanical state opens the possibility to study it from the perspective of quantum information theory, which studies the information content and correlations of quantum mechanical systems. Applying this framework to a trained TNML model can give us deep insights into the way the model processes information, and based on what criteria decisions are made.

\subsubsection{Schmidt decomposition}
An important concept of quantum information theory is the Schmidt decomposition. 
If $\mathcal{H}=\mathcal{H}_A\otimes\mathcal{H}_B$ is a bipartite Hilbert space with $\dim\mathcal{H}_A=n$ and $\dim\mathcal{H}_B=m$, then the Schmidt decomposition of a vector $\boldsymbol{w}\in\mathcal{H}$ is defined as
\begin{equation}
    \boldsymbol{w} = \sum_{i=1}^{r}\alpha_i\boldsymbol{u}_i\otimes \boldsymbol{v}_i
    \label{eq:schmidt}
\end{equation}
with orthonormal sets $\{\boldsymbol{u}_1,..,\boldsymbol{u}_r\}\in\mathcal{H_A}$ and $\{\boldsymbol{v}_1,..,\boldsymbol{v}_r\}\in\mathcal{H_B}$, and real, non-negative Schmidt coefficients $\alpha_i$. The Schmidt rank $r\leq\min\{n,m\}$ is the number of non-zero Schmidt coefficients. The Schmidt coefficients are unique up to reordering, and in the following we assume that they are ordered decreasingly. Note that this is essentially a reformulation of the singular value decomposition (SVD) in another context. For TN without loops, like MPS or TTN, the Schmidt decomposition can be computed efficiently through a series of local SVD and QR decompositions on the tensors~$\mathcal{T}$ (see e.g., \cite[Sec.~4.1.3]{Schollwoeck2011} for details). 

\subsubsection{Entanglement entropy}

Analogous to the Shannon entropy in classical information theory, the bipartite entanglement entropy, or von Neumann entropy, is the basic measure of information in a quantum mechanical system. It quantifies the entanglement over a \emph{bipartition} $A;B$ of a quantum mechanical system $\mathcal{H}$ into two complimentary subsystems $\mathcal{H}_A$ and $\mathcal{H}_B$ with $\mathcal{H}=\mathcal{H_A}\otimes\mathcal{H}_B$. It can be expressed in terms of the Schmidt coefficients $\alpha_i$ of the bipartition as
\begin{equation}
    H(A;B) = -\sum_i\alpha_i^2\log\alpha_i^2\, .
\end{equation}
Note that this is the Shannon entropy of the squared Schmidt coefficients, which define a probability distribution if the state is normalized. The entanglement entropy quantifies the amount of information that is shared by the two subsystems. In the context of machine learning, the weight tensor $W$ is defined on a product of Hilbert spaces $W\in\mathcal{H}=\mathcal{H}_1\otimes\cdots\otimes\mathcal{H}_n\otimes\mathcal{H}_l$, one for each feature and one for the label. This is due to the fact that the feature map $\Phi$ maps each feature $x_i$ to a local Hilbert space $\mathcal{H}_i$, and the additional index that corresponds to the output $f$ can also be interpreted to be a vector of a Hilbert space $\mathcal{H}_l$. This means that we can compute the entanglement entropy of any bipartition of features and the label efficiently through the Schmidt coefficients of the TN. In particular, the entropy of the bipartition of a single feature with the remaining system is of interest. This quantity tells us how relevant this feature is for the classification, as the entanglement entropy quantifies the amount of information shared between the feature and the rest of the model. Note that the quantity is computed from the model itself, without having to consider any data samples. It is therefore a measure of feature importance that is data independent and exact, i.e., without sampling error.

\subsubsection{Compression}

Another powerful property of TN is the ability to compress them in a controlled fashion, discarding the least relevant information first. To see how this is done, consider again the Schmidt decomposition (Eq.~\ref{eq:schmidt}). Now note that in a loop free TN, each link $v_m$ corresponds to a bipartition of the system, with a corresponding Schmidt decomposition. If the bond dimension $\chi$, i.e., the dimension of a link, is larger than the corresponding Schmidt rank $r$, then it can be \emph{truncated} without loosing any information, since the extra dimensions have Schmidt values that are equal to zero. When all bond dimensions are equal to the corresponding Schmidt ranks, the network is said to be of full rank. In the same spirit, the TN can be compressed further by discarding the smallest $k$ Schmidt values of a bipartition, s.t.
\begin{equation}
    \sum_{i=0}^{k-1}\alpha_{r-i}^2\leq\varepsilon,
    \label{eq:eps}
\end{equation}
where $\varepsilon>0$ is a tunable error threshold, and $k$ is chosen to be as large as possible while still fulfilling this condition. By repeating this procedure for all links $v_m$, always discarding the Schmidt values of the corresponding bipartition such that Eq.~\ref{eq:eps} is satisfied, the full network is compressed. This compression introduces an error in the representation of the full tensor, but since the Schmidt values correspond to singular values, we know that the error of this compression, with respect to the Frobenius norm, is given by the sum of the squared Schmidt-/ singular values that were discarded.

\section{Numerical experiments}

For numerical evaluations, we used the well established MSTAR dataset \cite{KeydelLeeMoore1996,RossEtAlBryant1998}.
The MSTAR (Moving and Stationary Target Acquisition and Recognition) dataset is a publicly available benchmark developed by the U.S. DARPA and AFRL for automatic target recognition using Synthetic Aperture Radar imagery. It contains high-resolution SAR images of military ground vehicles captured under varying depression angles and configurations. The dataset includes multiple target classes, such as tanks and armored personnel carriers, with labeled imagery for training and testing recognition algorithms.
It has been used in a wide range of applications (e.g., \cite{BlaschEtAlVelten2020,FeinAshleyEtAlBusart2023,SchumacherRosenbach2004,SinghSingh2023}).
Here, we employ the main $8$ target classes. The images were compressed to $32\times 32$ pixels, and the dataset was split randomly into a training set containing $70\%$ (6626) of the samples and a test set with the remaining $30\%$ (2840).

\subsection{Robustness against data poisoning}
\begin{figure}
    \centering
    \includegraphics[width=0.9\linewidth]{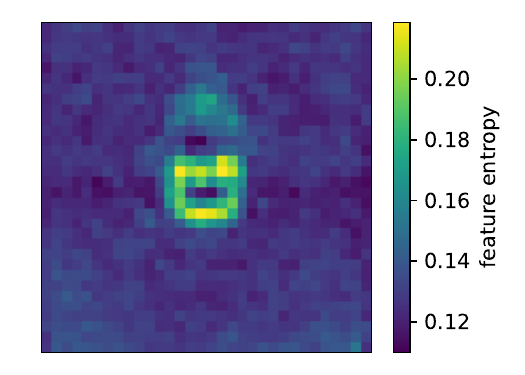}
    \caption{Feature entropies of the TTN model trained on the original MSTAR dataset.}
    \label{fig:entropy_clean}
\end{figure}

\begin{figure}
    \centering
    \includegraphics[width=0.9\linewidth]{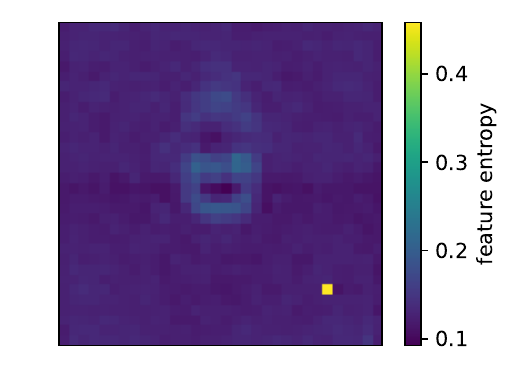}
    \caption{Feature entropies of the TTN model trained on a poisoned dataset where one single pixel in the background was correlated to the target class.}
    \label{fig:entropy_pixel}
\end{figure}

\begin{figure}
    \centering
    \includegraphics[width=\linewidth]{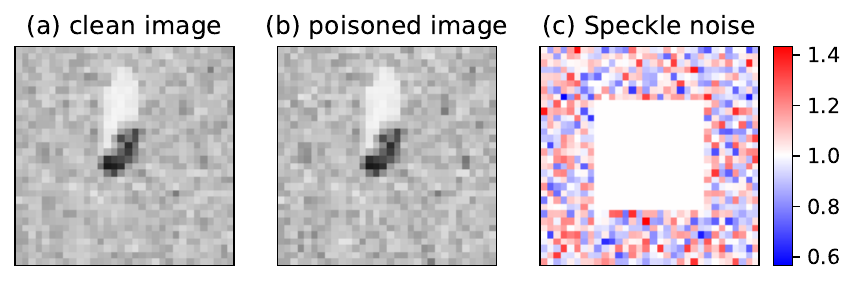}
    \caption{Illustration of the noise that was applied to the background. The difference between the original image (a) and the poisoned image (b) is barely visible. c) The speckle noise that was injected to the background region $\mathcal{D}$. The region in the center, where the target is located, is left unchanged. This specific noise pattern was applied to all samples of this class, and a different pattern to the other classes, respectively, but all generated from the same normal distribution.}
    \label{fig:poisoned_image}
\end{figure}

\begin{figure}[t!]
    \centering
    \includegraphics[width=0.9\linewidth]{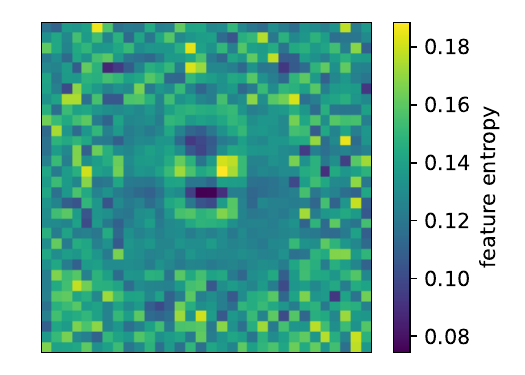}
    \caption{Feature entropies of the model TTN trained on a poisoned dataset where the background was strongly correlated to the target class.}
    \label{fig:entropy_background}
\end{figure}

In the MSTAR dataset, all targets are located at the center of the images and are surrounded by a background of clutter. A common pitfall of this dataset is that machine learning models learn to classify using information in the background, rather than the targets themselves. This will lead to poor performance on real data, where the background might be different. With TN models, we can easily assert that the model learned to use information of the targets by looking at the feature entropies of the model. In \textbf{Figure~\ref{fig:entropy_clean}}, one can see the feature entropies of a TTN model that was trained on the MSTAR dataset, reaching a test accuracy of $99.05\%$. In the area where the target is located, the feature entropies are significantly higher than in the surrounding area. Since the feature entropies are a measure of a feature's importance for the classification, this tells us that the model learned to use the information of the targets, while mostly neglecting the background.
To demonstrate the effectiveness of this method, we now artificially correlate a single pixel (feature) located in the background with the class for all samples. If $k$ is the index of the pixel we modify, then the feature $x_{i,k}$ of sample $i$ is replaced by
\begin{equation}
    \bar{x}_{i,k} = \frac{8 - y_i}{10} \cdot X\:,
\end{equation}
where $X$ represents multiplicative, Gaussian distributed noise $X\sim\mathcal{N}(1,\,10^{-4})$ and $y_i=\{0,1,...,7\}$ is the numerical value assigned to the class of sample $i$. The modification was chosen such that the modified pixel blends well with the clutter background, making it difficult to detect the change by simply looking at the data. 
The model trained on this data reaches $99.3\%$ test accuracy on the modified test data, but only $92.68\%$ on the original (clean) test data. This suggests that the model is relying on the information encoded in the modified pixel, but without access to the clean data this cannot be tested. However, by looking at the feature entropies again (\textbf{Figure~\ref{fig:entropy_pixel}}), we can immediately see that something is wrong, as the feature entropy of the modified pixel is by far the largest.
Instead of replacing a single feature, we can also correlate the whole background to the target class by injecting multiplicative Gaussian noise (speckle) that depends on the target class. The noise is only applied to the clutter background $\mathcal{D}$, to not affect the targets and make the modification less obvious (see \textbf{Figure~\ref{fig:poisoned_image}}). The modified features $\bar{x}_{i,k}$ are given by
\begin{equation}
    \bar{x}_{i,k} = \begin{cases}
        x_{i,k} \cdot n_{y_i,k}, \quad &k\in\mathcal{D}\\
        x_{i,k}, &\textrm{else}
    \end{cases} 
\end{equation}
with $n_{y,k}\sim \mathcal{N}(1,\,0.02)$. The TTN classifier trained on this data achieved a test accuracy of $99.86\%$ on the poisoned data, and $82.18\%$ on the clean test data. In \textbf{Figure~\ref{fig:entropy_background}}, one can see the feature entropies of this model. Clearly, the model is relying heavily on the modified background, which explains the significant drop in accuracy on the clean test data, where the patterns in the background that the model learned are not present. 

\subsection{Model compression}

\begin{figure}
    \centering
    \includegraphics[width=\linewidth]{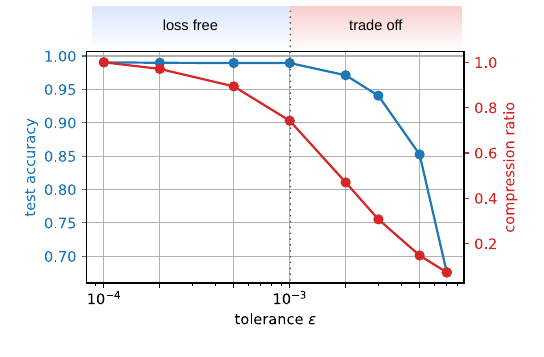}
    \caption{Comparison of the number of parameters in the model and the test accuracy of the compressed model with increasing compression threshold $\varepsilon$.}
    \label{fig:compression}
\end{figure}

To demonstrate the power of tensor network compression in the context of machine learning, we compress the model trained on the MSTAR dataset with gradually increasing error threshold $\varepsilon$, which defines the maximal sum of squared singular values that are discarded (Eq.~(\ref{eq:eps})). In \textbf{Figure~\ref{fig:compression}}, one can see how the number of parameters of the model, which is proportional to the inference time, and the test accuracy change as the compression tolerance $\varepsilon$ is increased and the model gets compressed more strongly. This particular model can be compressed up to a threshold of $\varepsilon=10^{-3}$ without a loss of accuracy, while the number of parameters is decreasing. In this loss-free domain, the model can be compressed up to a compression ratio 
$$r=\frac{\text{compressed size}}{\text{uncompressed size}}$$ of about $75\%$. With higher tolerances, accuracy gradually decreases along with the number of parameters. For $r\approx47\%$, the test accuracy is still at $97.11\%$. Since inference time is proportional to the number of parameters, there is a trade-off between accuracy and inference time/ model size in this domain.

Similar behavior can be observed for all TNML learning models, although the concrete relationship between the compression tolerance, the accuracy and the number of parameters depends on many factors such as the quality of the data or the model's ability to generalize.

\section{Summary}

This study investigated the application of tensor networks in radar technologies, particularly for SAR object classification.
The inherent challenges posed by the vast dynamic range and noise in SAR imagery necessitated a strong emphasis on robustness. 
We evaluated the resilience of tensor networks to data poisoning, including the presence of poisonous pixels, through the MSTAR SAR classification task. This task was particularly suitable due to its moderate dataset size and substantial target variability, allowing for a comprehensive assessment of model reduction and robustness.
The findings revealed that tensor networks achieved significant model reduction while maintaining high classification accuracy, making them ideal for deployment in resource-constrained environments. Furthermore, this research emphasized the robustness and model reduction capabilities of tensor networks in object classification, indicating that they were well-suited to tackle the dual challenges of enhancing robustness and achieving model efficiency. Overall, the results contributed valuable insights to the ongoing discourse in radar applications and deep learning methodologies, since model reduction and robustness are becoming more and more important as models are deployed to an ever wider set of applications.

\section{Outlook}

The results presented here have important practical consequences for the enhancement of object detection pipelines. First, the ability to tune models by balancing computing power, desired inference time, and model performance facilitates context-dependent object identification. That is, models can be tweaked depending on the importance of high accuracy and whether or not longer computation times are acceptable. Second, the ability to identify data poisoning lends itself to a degree of explainability, which is also important in practice. Both of these points can be seen, for example, in a military setting, where being able to adapt a classification model to the context could result in operational advantages. On top of this, having the ability to identify adversarial attacks or when a mission-critical model is making ill-informed decisions could assist a human operator in making important decisions quickly and confidently. 

While the data used here demonstrate the potential of the proposed approach in the defense domain, the approach proposed here is certainly not restricted to this setting. An example target classification pipeline could look as follows: Spaceborne radar systems collect data over specific areas of interest. A TNML model would then be selected based on the operational context, and the resulting classifications, combined with feature importance information, this would give additional information to a human actor to aid their decision-making. 
In general, this work follows the trend of incorporating quantum and quantum-inspired algorithms in radar, and further work could include testing data sets from different domains and comparing results to traditional machine learning approaches.

\section{Acknowledgements}
This work is part of the QUA-SAR project by the DLR Quantum Computing Initiative (DLR QCI) and is funded by the Federal Ministry of Research, Technology and Space (BMFTR). \url{https://qci.dlr.de/qua-sar}
\newpage
\printbibliography

\end{document}